\documentclass{report}

\usepackage{cite}
\usepackage{amsmath,amssymb,amsfonts}
\usepackage{algorithmic}
\usepackage{graphicx}
\usepackage{textcomp}
\usepackage{float}
\usepackage{url}
\usepackage{xcolor}
\usepackage{subcaption}

\title{
\begin{center}
        \includegraphics[width=0.9\textwidth]{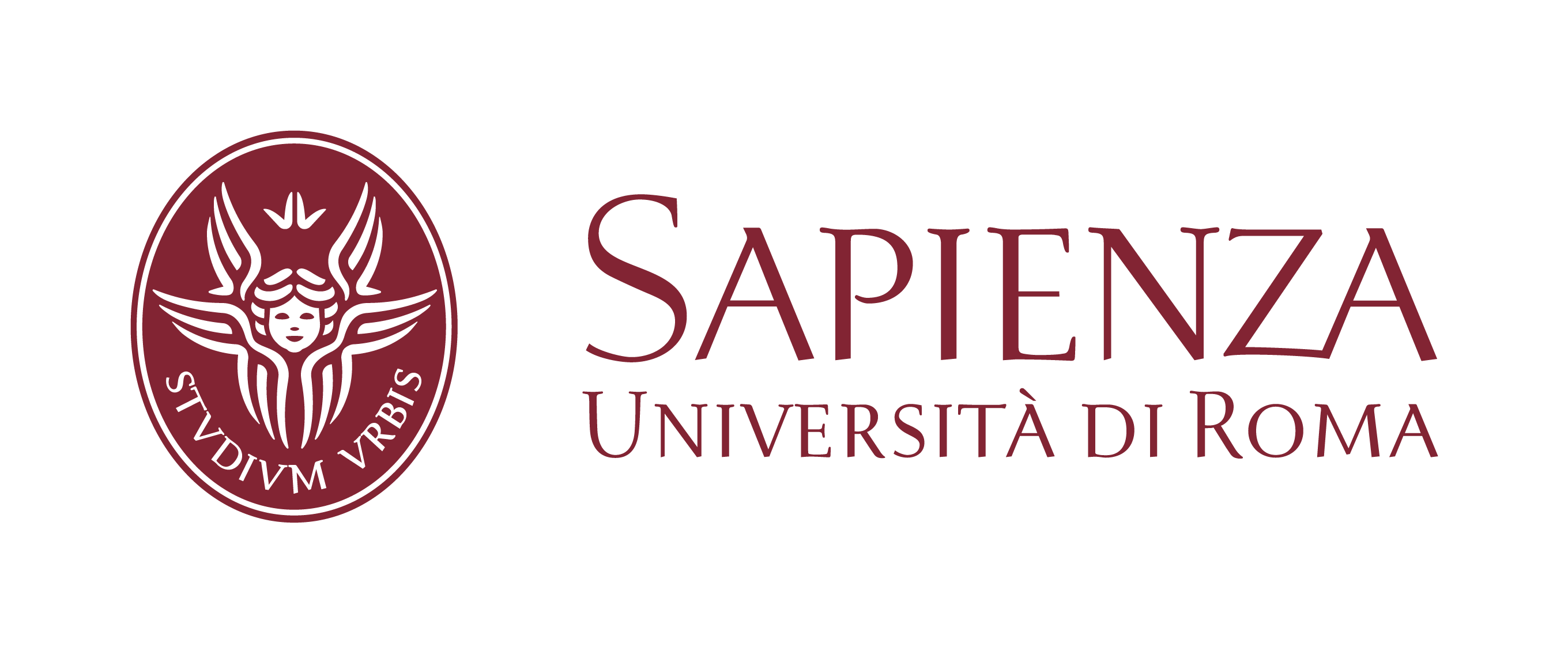}
    \end{center}
A Proof of Concept for a Digital Twin of an Ultrasonic Fermentation System
}

\author{
Francesco Saverio Sconocchia Pisoni\\
\small Sapienza University of Rome\\
\small \texttt{sconocchiapisoni.1889241@studenti.uniroma1.it}
\and
Andrea Vitaletti\\
\small Sapienza University of Rome\\
\small \texttt{vitaletti@diag.uniroma1.it}
\and
Davide Appolloni\\
\small University of Rome Tor Vergata\\
\small \texttt{davide.appolloni@students.uniroma2.eu}
\and
Federico Ortenzi\\
\small University of Rome Tor Vergata / Yeastime Srl\\
\small \texttt{federico.ortenzi@yeastime.com}
\and
Blasco Morozzo della Rocca\\
\small University of Rome Tor Vergata\\
\small \texttt{blasco.morozzo.della.rocca@uniroma2.it}
\and
Mariano José Guillén\\
\small Yeastime Srl\\
\small \texttt{guillen.mj98@gmail.com}
\and
Alessandro Contaldo\\
\small Yeastime Srl\\
\small \texttt{alessandro.contaldo@yeastime.com}
}
\date{}
\begin{document}

\maketitle

\begin{abstract}
This paper presents the design and implementation of a proof of concept digital twin for an innovative ultrasonic-enhanced beer-fermentation system, developed to enable intelligent monitoring, prediction, and actuation in  yeast-growth environments. A traditional fermentation tank is equipped with a piezoelectric transducer able to irradiate the tank with ultrasonic waves, providing an external abiotic stimulus to enhance the growth of yeast and accelerate the fermentation process. At its core, the digital twin incorporates a predictive model that estimates yeast’s culture density over time based on the surrounding environmental conditions. To this end, we implement, tailor and extend the model proposed in \cite{Palacios_2014}, allowing us to effectively handle the limited number of available training samples by using temperature, ultrasonic frequency, and duty cycle as inputs. The  results obtained along with the assessment of model performance demonstrate the feasibility of the proposed approach.

\end{abstract}

\tableofcontents


\chapter{Introduction}

\begin{figure*} [ht]
    \centering
    \includegraphics[width=\linewidth]{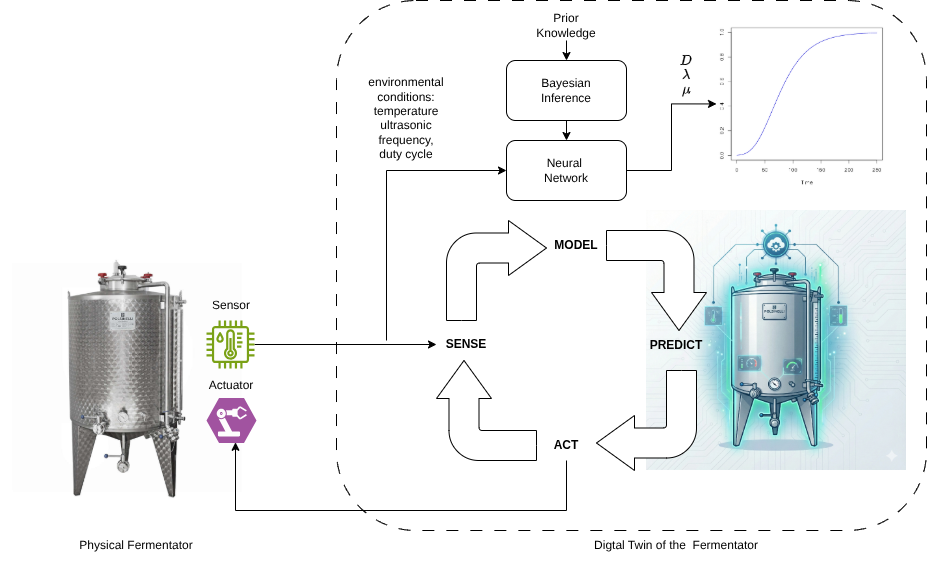}
    \caption{A schematic overview of the proposed solution. On the left there is the physical fermentation tank, connected via the sensors to the Digital Twin, whose scheme is reported on the right. Inverse communication is established via the Act flow and implemented by actuators. The environmental parameters, temperature, ultrasonic frequency, and duty cycle,  serve as inputs to a neural network that infers the parameters $D, \lambda, \mu$ of the Gompertz model used to estimate the growth of the yield. To address the limited size of the available dataset, prior knowledge is incorporated into the neural network through Bayesian Inference, allowing the model to remain robust despite sparse observations. }
    \label{fig:DTarch}
\end{figure*}

A digital twin \cite{10818423} is a virtual representation of a physical system that stays continuously synchronized with it through real-time data, enabling real-time monitoring, predictive simulation, and closed-loop control. In other words, it is a live, data-driven model of a real object, process, or environment. Specifically, sensors on the physical system send continuous data to the digital model, keeping the digital twin up-to-date with what is happening in reality. This data also trains and feeds a model, which is the core component to run simulations to predict performance, failures, and behavior under different conditions. This enables testing different scenarios, including most problematic ones, without putting at risk the real system. The results of the simulations inform decision-makers and guide the actions into the physical system, enabling adaptive control, automated tuning, and optimization of operation. In summary, a typical digital twin (see figure \ref{fig:DTarch}) forms an iterative loop,  continuously improving with more data, encompassing the following tasks: sense, model, predict, act. This enables faster experimentation rounds (especially when physical tests are slow or expensive), cost reduction, enhanced reliability, intelligent automation, and continuous improvement through data-driven feedback.

In fermentation environments, experiments are slow and resource-intensive; therefore, a digital twin accelerates exploration by simulating environmental conditions such as temperature, glucose and pH, and predicting yeast growth without requiring repeated physical trials or highly complex procedures. Estimates by Yeastime\footnote{\url{https://yeastime.com/}} indicate that a 30\% reduction in fermentation time in craft breweries could lead to cost savings of approximately 10–15\%, with clear economic benefits.

In this paper, we focus on the innovative application of ultrasound technology to the fermentation process, with the aim of developing a PoC of a digital twin to evaluate its impact and potential benefits. 

\section{Background: ultrasound in fermentation processes}

Fermentation processes are biologically mediated chemical transformations, crucial to many aspects of human food and beverage. They typically consist of the metabolic process where yeast and/or bacteria anaerobically convert organic substrates (mainly carbohydrate sources, such as glucose, maltose etc) into various products, such as organic acids, ethanol and gases (i.e. carbon dioxide) without the use of light as energy source \cite{DUSSAP2017263}.
As a concise yet illustrative list of traditional applications, we may cite wine, beer, coffee, bread, kombucha, kefir, and various dairy products. More recent developments include the use of fermentation processes to exploit cells as production platforms for targeted high-value compounds, a strategy that underpins the emerging field of precision fermentation. 
The usage of fermentation by humans dates back thousands of years and although being a very artisanal and traditional practice, it has focused the interest of research and industrial innovation in the last decades. Beer brewing is a prominent industrial fermentation process; beer remains one of the world’s most consumed beverages and its market continues to expand \cite{colen2016economic}. The modern brewing production pipeline is the result of diverse optimization strategies, from the malting phase \cite{munoz2013malting}  to the bottling one \cite{nieto2024effects}. Nevertheless, the imperative to improve the sustainability of industrial processes \cite{olajire2020brewing} has driven substantial research into optimizing yeast fermentation.
One very recent approach to optimization has spawned from the development of mechanosensing biology, the study of the capacity of living organisms to sense and react to mechanical cues, including pressure waves. Thus sound, i.e. a mechanical wave propagating via local pressure variation, is now introduced as an additional abiotic parameter influencing the organism growth and metabolic characteristics. This is not limited to the range of human audible sounds (i.e. 2 to 20 KHz) but also infra and ultrasounds, for which frequencies are below or above the human audible frequency range, respectively. In more detail three quantities define sound: intensity, frequency and phase. Pivotal dimensions are pressure, which corresponds to sound amplitude and is expressed as Sound Pressure Level (SPL) and frequency expressed in Hertz (Hz). \cite{leighton2007ultrasound}.\\ 

Mechanosensitive (MS) channels are specialized membrane proteins that function as biological force transducers\cite{Marshall2012-or}. They sense changes in membrane tension and physical stress, converting these mechanical stimuli into electrochemical signals by opening a pore to allow ion flux \cite{Jin2020-wp}.
In bacteria, MS channels primarily act as "safety valves" to prevent cell lysis during hypoosmotic shock \cite{Martinac2011-id}. When a bacterium is suddenly exposed to a low-solute environment, water rushes into the cell, increasing turgor pressure. MS channels open to release cytoplasmic solutes, relieving this pressure. The prototypical protein is MscL (Mechano-Sensitive Channel of Large conductance) \cite{Rajeshwar_T2021-kj}, as it is the most well-studied MS channel. It requires significant membrane tension to open and serves as the last line of defense against cell bursting. McsL are often complemented in bacterial cells by MscS (Mechano-Sensitive Channel of Small Conductance): this family is more diverse and opens at lower tension levels than MscL, providing a more nuanced response to osmotic changes\cite{Cox2019-ei}. 
In eukaryotes, MS channels are involved in sophisticated sensory functions, including touch, hearing, proprioception, and blood pressure regulation. Unlike the "emergency valve" role in bacteria, eukaryotic channels are often integrated into complex signaling pathways. Some examples of eukaryotic MS channels are: Piezo Family (Piezo1 and Piezo2): these are large, trimeric proteins with a unique "propeller" shape \cite{piezo_first}. They are essential for touch sensation and sensing fluid shear stress in blood vessels, and their study was recognized with the Nobel Prize in Physiology or Medicine in 2021; K2P Channels (e.g., TREK-1\cite{WIEDMANN2021126}): These potassium channels are sensitive to membrane stretch and help regulate the resting membrane potential in response to mechanical deformation; TRP Channels: Various members of the Transient Receptor Potential family (like TRPV4 \cite{ji_trp4}) are polymodal, responding to both temperature and mechanical force \cite{LIU201522}.
\\
Ultrasound technology, at low power and low frequency, has emerged as a promising tool to enhance beer fermentation by accelerating yeast activtiy and fermentation kinetics, as demonstrated by Kwon et al.~\cite{kwon2020ultrasound}. The application of ultrasound to barley during fermentation improves sugar availability and fermentation efficiency. In their work, Alonso et al. \cite{alonso2021ultrasonication} irradiated beer during primary fermentation over a period of four days at 40 kHz, varying power levels, and irradiation durations. Under optimal conditions of 160 W power and 12 hours of treatment, ethanol production was enhanced by 13.18\%.
However, an adverse effect was observed when the power was increased to 200 W, indicating the existence of an optimal operational window. In more detail, these reports underline that a power dependent ultrasound effect influences treatment output. A key phenomenon that ultrasounds can trigger is acoustic cavitation. This emerges when static pressure of a liquid reduces below the liquid vapour pressure, forming vapour filled cavities. When higher pressure hits the liquid, these cavities or bubbles exceed the attractive forces and collapse, generating shock waves \cite{franc2005fundamentals}. The effect on the motion of cavities may be violent, or gentler, allowing to distinguish between stable or transient cavitation. At high ultrasound amplitude, transient phenomena emerge with bubble collapse impacting cell viability and negatively affecting the target process. On the other hand, low amplitude induces stable cavities, bubbles, that oscillate around some equilibrium size leading to microstreaming and mixing events around the cell \cite{neppiras1980acoustic}.
Ojha et al. \cite{OJHA2017410} provided a broader analysis of ultrasound technology applications in food fermentation. They report that frequencies in the range of 20–50 kHz improve mass transfer and cell permeability, leading to enhanced process efficiency and production rates Studies on convection flow during fermentation ~\cite{lin2013convection} reveal the role of ultrasound in optimizing mass transfer throughout the cell membrane. This lead to increased nutrient permeability and decreasing the concentration of pericellular feedback inhibitory metabolites, which in turn leads to enhanced metabolic processes and an increase in biomass \cite{pawar2020role}. Moreover, controlled ultrasound cycles have been demonstrated to improve solid-liquid interface increasing gas transfer, to disassemble cell clusters \cite{schlafer2002ultrasound} . 
Comprehensive reviews highlight the ability of low-power ultrasound to reduce fermentation time, improve ester profiles, and improve sensory quality in beer~\cite{zhang2023ultrasound_review}. 
\\

What is limiting the spread of ultrasound-based systems into industrial scale facilities is the lack of integrated bioreactors where to irradiate reproducible ultrasound cycles. A bioreactor, i.e. a fermenter, refers to any vessel used to carry out a biochemical reaction where a substrate to product conversion occurs. Microbes are grown in these dedicated systems, while photosynthetic organisms, as plant cells, microalgae or cyanobacteria require light permeable systems to grow, defined photobioreactors (PBRs). We find different kinds of closed systems as tubular, horizontal, vertical PBRs \cite{tredici2009photobioreactors}. These are controlled systems that allow the setting of the optimal parameters required to cultivate a defined strain or to produce a target product, as pH, temperature, nutrient levels or agitation \cite{liden2002understanding} . Bioreactors allow the fine modulation of key operational factors: at first to maintain monoseptic operation, oxygen or nutrient supply \cite{gopalakrishnan2024industrial}  and carbon dioxide removal, heat transfer to control temperature and shear stress control to avoid damage to the cell culture, PBRs also allow the modulation of lighting conditions (frequency and light:dark cycle) \cite{carvalho2011light} . The operative parameters of the bioreactor are set so as to fulfill the culture’s requirements and to set up cost-effective processes that minimize resource loss. Since each culture has diverse metabolic needs and culture’s biomass and high value product yields are strongly influenced by their interaction within the micro-environment \cite{liden2002understanding} , the challenge is to find, implement and apply dedicated parameters to optimise the process \cite{delavar2022advanced} . Culture mixing is of great interest to guarantee an homogeneous environment, it increases access to nutrients for heterotrophs and light for phototrophs. To this aim batch reactors are usually equipped with rotors, stirrers or with airlift or air insufflation technologies \cite{tredici2009photobioreactors}. These approaches can favor mass transfer and increase liquid phase exchanges, but are hampered by cellular stress induction \cite{berry2016characterisation}. Shear stress, induced by rotors or bubble flows, is a central factor to consider in industrial processes, so much that microalgae are classified for their shear stress sensitivity \cite{wang2018effects}. Hence, the successful integration of pressure wave sources into bioreactors must consider the working parameters of the target process. 
\\

In this scenario, the landmark paper of Chisti \cite{chisti2003sonobioreactors}  envisages diverse configurations of sonobioreactors. The implementation of a defined experimental setting allows reproducible bio-effects to emerge. An application of this study is proposed and presented in the novel sono-photobioreactor ~\cite{ortenzi2025sono}, system where ultrasound technology is exploited to optimise microalgal-based processes, at laboratory scale. Industrial applications such as those presented by Yeastime, demonstrate the practical benefits of ultrasound technology in reducing the duration of fermentation by up to 30\%, contributing to increasing the sustainability of the overall brewing processes~\cite{yeastime2024ultrasound}. \\

\emph{Collectively, these references highlight the significant potential of ultrasound integration into novel sonobioreactor in innovating traditional beer fermentation by improving efficiency and product quality, and suggest to initially focus on frequency in the range 20–50 kHz}.

\section{Related Work}
Digital twin technology is increasingly applied to biological systems for enhanced modeling, prediction, and control. Kim and Lee~\cite{kim2024digital} provide a foundational review of digital twins in biological contexts, setting the stage for their applications in biotechnology. Nguyen and Smith~\cite{nguyen2025digital} extend this approach to healthcare, discussing multi-scale digital twins that include gene-regulatory and cellular network models, relevant to fermentation microbial systems. The FermenTwin project~\cite{fermentwin2025} exemplifies digital twin implementation specifically for fermentation, aiming to replicate microbial dynamics and process parameters in real time. Wong et al.~\cite{wong2025optimization} demonstrate digital twins integrated with AI for bioprocess optimization and scale-up in intelligent manufacturing settings. Li and Zhao~\cite{li2025digital} focus on applying digital twins for prediction and control in food fermentation, showing improved process reliability and product quality. Together, these works highlight the transformative impact of digital twins in precision and efficiency of biological and fermentation process engineering.
To the best of the authors' knowledge, this is the first report of digital twin tailored on ultrasound-treated fermentation process.

\chapter{Sense}
The physical fermenter is equipped with sensors to monitor the ultrasound frequency applied to the medium, its voltage level, and the temperature of the fermenting mixture inside the bioreactor.

\section{Ultrasonic Sensor} A piezoelectric disk is used as a passive sensor to detect ultrasonic vibrations transmitted into the medium. Although its nominal bandwidth is limited to approximately 80 Hz–15 kHz, empirical tests using an oscilloscope showed that it still produces measurable responses up to about 40 kHz, despite significant attenuation and noticeable non-linear behavior. Although a more complex and expensive sensor could easily cover the required frequency range, we consider the piezoelectric disk appropriate for our initial experimentation.



\section{Temperature Sensing} A Dallas immersion probe (1-wire protocol) provides 12-bit measurements with 0.0625$^\circ$C resolution. A custom driver was implemented using ESP32 RMT timing primitives. It is important to note that while direct immersion offers excellent temperature accuracy, it requires sterilization and may still influence microbial growth by providing an adhesion site for the colonies under study. A more reliable, though more complex, approach using externally mounted probes or infrared thermal sensing is left for future developments of the system.

\chapter{Dataset}
The training of the model to perform predictions requires a proper dataset that is enriched by the acquisition of real-time data during the operation. In biological experiments, datasets are typically obtained through manual sampling performed by biologists. Such procedures are both time-consuming and costly, resulting in datasets that are relatively limited in size. In our case we have $\approx$100 points characterized by: ultrasound duty cycle, irradiation frequency, irradiation duration,  temperature, initial bacterial concentration, and final bacterial concentration.
Ultrasound duty cycle, irradiation frequency, and temperature values are normalized to lie between 0 and 1, preventing any one parameter from dominating the analysis. Points with the same duty cycle, irradiation frequency, temperature, and initial bacterial concentration belong to the same biological growth (see section~\ref{sec:predict}). Note that biological growth produces a variation in temperature that should be controlled by proper actuators.

\chapter{Model} 

The literature on biological models is vast \cite{muller2015methods}. In this work, we are interested in considering a simple model that allows us to estimate the growth curve for yeast populations under different environmental conditions. Similar models are widely applied to both bacteria and yeasts. We initially focus on empirical sigmoidal models such as Gompertz or Weibull, used to fit growth curves or product formation for both bacterial and yeast processes \cite{García02042024, 10.1111/lam.12314}. Specifically, we refer to the work in \cite{microorganisms12071449}, which evaluates the growth of the yeast \textit{Saccharomyces} \textit{cerevisiae}, the model organism used in the fermentation processes of our experiments, using predictive growth models (Buchanan’s, modified Gompertz, and Baranyi–Roberts) under different initial glucose concentrations. We analyze how various environmental factors, such as temperature, acidity, and sugar concentration, affect yeast growth, using the methodologies presented in \cite{Palacios_2014}.

The core of the digital twin is a predictive model that estimates yeast density over time based on the surrounding environmental conditions. To achieve this, we adopt the widely used Gompertz model \cite{10.1371/journal.pone.0178691} described by the following equation. 

\begin{equation}\label{eq:gompertz_fun}
\begin{split}
E[N_t | N_0, D, \mu, \lambda] = g(t, N_0, D, \mu, \lambda) \\
\text{where }\\ 
g(t, N_0, D, \mu, \lambda) = N_0 + D \exp\left(-\exp\left(1 + \frac{\mu e(\lambda - t)}{D}\right)\right)
\end{split}
\end{equation}
The Gompertz model parameters in Equation~\ref{eq:gompertz_fun} are defined as follows:
\begin{itemize}
    \item $N_t$: yeast density at time $t$, available in the dataset;
    \item $N_0$: initial yeast density, available in the dataset;
    \item $D$: total growth amplitude, representing the difference between the maximum and initial yeast density;
    \item $\mu$: specific growth rate at the inflection point;
    \item $\lambda$: characteristic time corresponding to the inflection point of the growth curve.
\end{itemize}

The model parameters $D$, $\mu$ and $\lambda$, are inferred by a single-hidden-layer neural network with sigmoid activations, which takes the temperature of the medium, the ultrasonic frequency, and the duty cycle of the irradiating transducer as inputs. 
The use of sigmoid activation is well justified, since it is typical of bacterial growth and yeast to exhibit a time-dependent behavior characterized by an initial stabilization phase (lag phase), followed by an exponential growth phase (log phase), and finally a stationary phase.
In the latter, the number of living cells equals the number of dead cells, leading to stabilization at a characteristic bacterial density. A function that closely resembles this type of behavior in shape is the sigmoid function, which is therefore particularly suitable to be employed as the
activation function within the neural network.


Given the limited amount of data available to train the model, the employment of classical training methods might be ineffective. For this reason, we adopted the methodology proposed in \cite{Palacios_2014}, in which Bayesian inference is used to estimate the neural-network–based growth model while incorporating prior biological knowledge. Prior distributions, namely information about the biological phenomenon known before looking at the specific dataset, are assigned to the network weights to regularize learning and reduce the reliance on larger datasets. This Bayesian formulation enables the model to constrain parameters to biologically plausible ranges.
In our case, we were forced to use highly informative priors given the limited dataset size; for larger datasets, a weakly informative and simple prior on the weights, such as $\mathcal{N}(0, 100)$, is recommended, as also stated in \cite{Palacios_2014}.





The whole approach can be summarized as follows: 
\begin{enumerate}
\item The process begins by assigning probability distributions to the neural network weights to incorporate prior knowledge
\item Random Walk Metropolis is then used to explore the posterior landscape. This landscape represents the joint probability of the parameters given the observed data (likelihood) and the prior. The algorithm generates multiple plausible states (i.e., a specific set of weights). To escape from local maxima, Metropolis-Hastings Acceptance Criterion is used. 
\item Each state instantiates a distinct neural network $i$. This network takes in input the duty cycle, irradiation frequency and temperature 
 of a specific point in the dataset and produces as output the three Gompertz parameters:  $D_{i}$ (amplitude), $\mu_{i}$ (growth rate), and $\lambda_{i}$ (lag time).  
\item The final prediction is performed by equation~\ref{eq:gompertz_fun} where the parameters $D,\mu,\lambda$ are computed by averaging the Gompertz parameters of all the neural networks computed in the above step. 
\end{enumerate}

\chapter{Predict: Model Performance}\label{sec:predict}

To assess model performance, the dataset was split into a training set (67\% of the data) and a test set (33\% of the data), using group 5-fold cross-validation. 
As described in~\cite{Palacios_2014}, data points belonging to the same biological growth curve must be treated as a group, and consequently cannot be split between the training and test sets.
A biological growth was uniquely identified by the combination of three experimental parameters: duty cycle, frequency, and initial yeast density.
These groups were then distributed across $k=5$ folds using a round-robin allocation strategy, yielding five distinct partitions of the dataset into respective training and test sets.
Cross-validation metrics were obtained by averaging the performance measures across all five folds.
The model was trained using 100,000 burn-in iterations, followed by 300,000 iterations for posterior distribution sampling, with a thinning rate of 500, yielding 600 samples for predictions. To assess model performance we considered the following metrics: 

\begin{itemize}
    \item \textbf{Mean Absolute Percentage Error (MAPE)}: the average error between each predicted value and its corresponding observed value in the test set, calculated using the absolute difference to prevent positive and negative errors from canceling out.
    \item \textbf{Median Absolute Percentage Error}: the median error, providing insight into whether performance is degraded by a few large outliers.
    \item \textbf{Mean Square Error (MSE)}: a performance metric enabling comparison with the paper~\cite{Palacios_2014} results. The MSE is computed as:
    \begin{equation}
        \text{MSE} = \frac{1}{n}\sum_{i=1}^{n}(y_i - \hat{y}_i)^2
    \end{equation}
    where $y_i$ represents the observed values, $\hat{y}_i$ the predicted values, and $n$ the number of test samples.
\end{itemize}

\begin{figure*}[htb]
    \centering
    \includegraphics[width=1\linewidth]{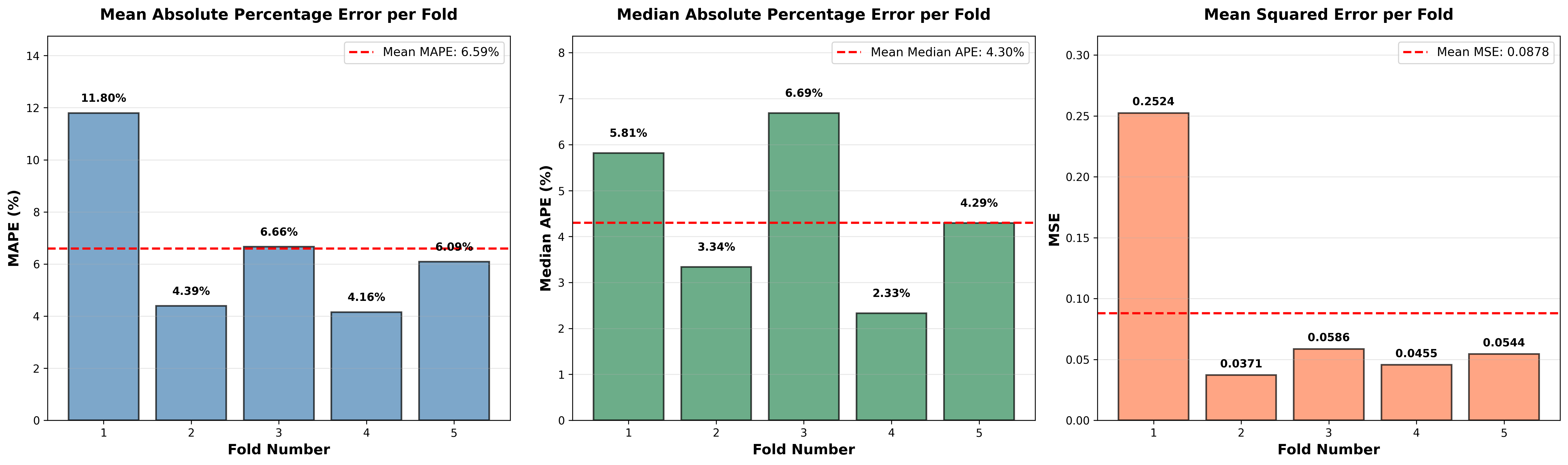}
    \caption{Distribution of performance metrics across folds. In the left panel Mean Absolute Percentage Error is displayed, in the central one the Mean Absolute Percentage Error, and on the right the Mean Squared Error.}
    \label{fig:metrics_fold}
\end{figure*}

The model achieves an MAPE of 6.59\%, a median absolute percentage error of 4.30\%, and an MSE of 0.0878 (see Figure~\ref{fig:metrics_fold}).

\begin{figure*}[h]

\begin{subfigure}{0.5\textwidth}
\includegraphics[width=0.9\linewidth]{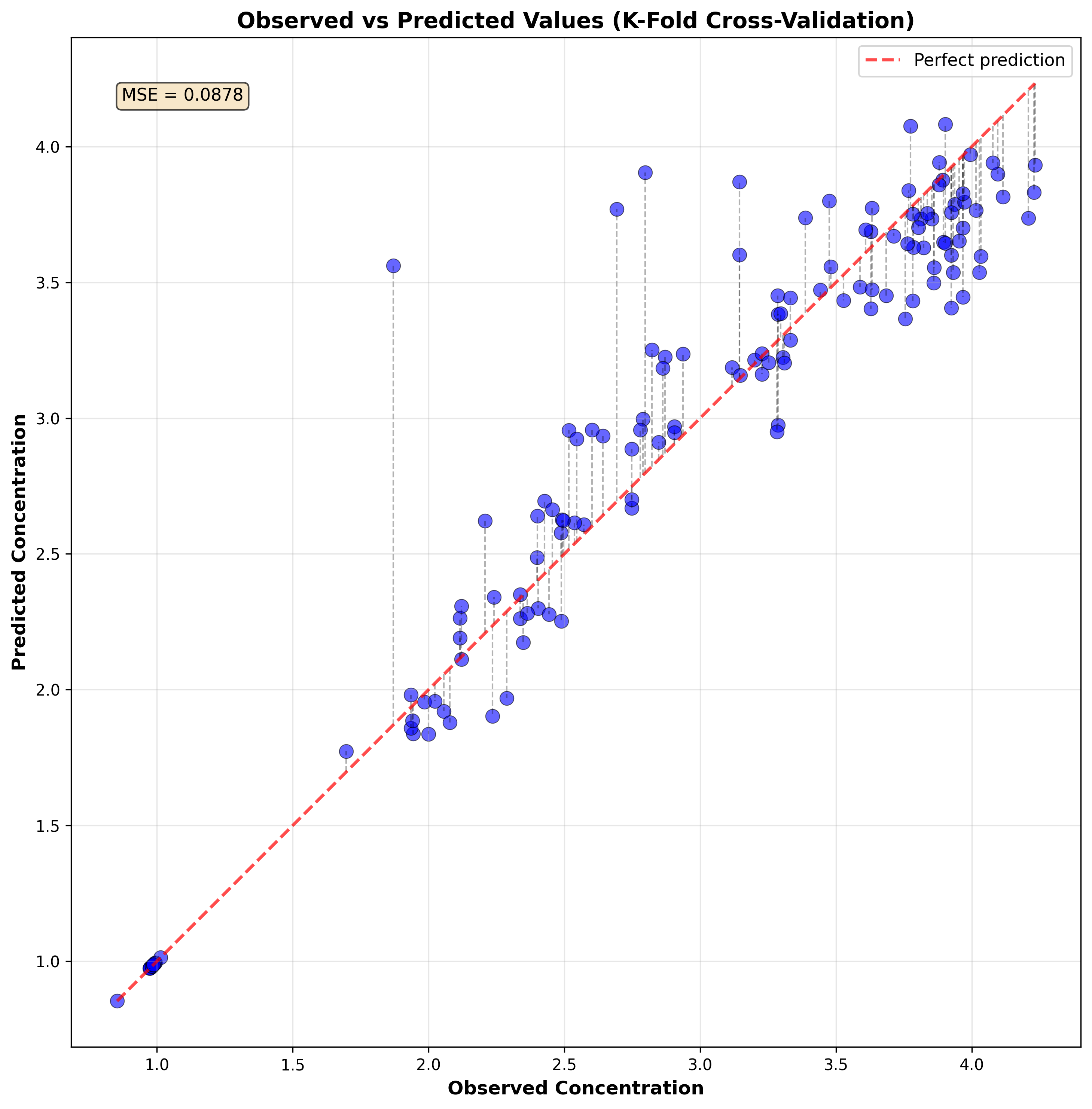}
 \caption{The black dashed lines indicate the error magnitude for each prediction.}
    \label{fig:errors}
\end{subfigure}
\begin{subfigure}{0.5\textwidth}
\includegraphics[width=0.9\linewidth]{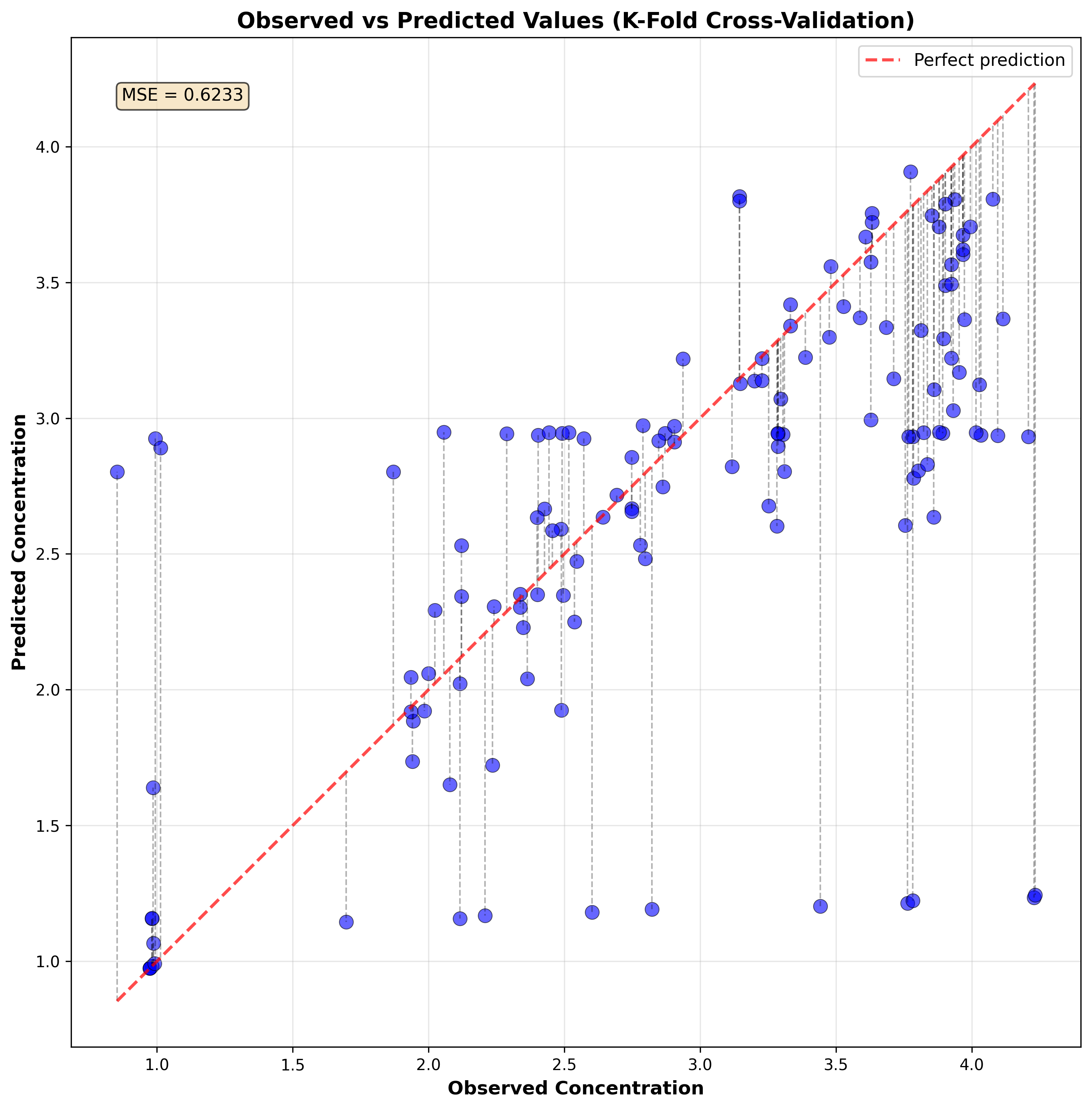}
\caption{Results using \cite{Palacios_2014} non-informative priors. The increased scatter compared to figure~\ref{fig:errors} demonstrates the importance of priors for limited datasets.}
    \label{fig:errors_no_inf}
\end{subfigure}

\caption{Predicted vs observed optical density at 600 nm values across all test points in the 5-fold cross-validation.}
\label{fig:image2}
\end{figure*}

This MSE is approximately 88 times larger than the value reported in \cite{Palacios_2014}. Nevertheless, a visual inspection of the predicted versus observed yeast concentration, measured through optical density (Figure \ref{fig:errors}), shows good qualitative agreement. This provides an initial validation of the model, indicating that it can predict yeast growth with reasonable accuracy, as further confirmed in Figure \ref{fig:growth_curve_comparison}.

\begin{figure*}[h]

\begin{subfigure}{0.5\textwidth}
\includegraphics[width=0.9\linewidth]{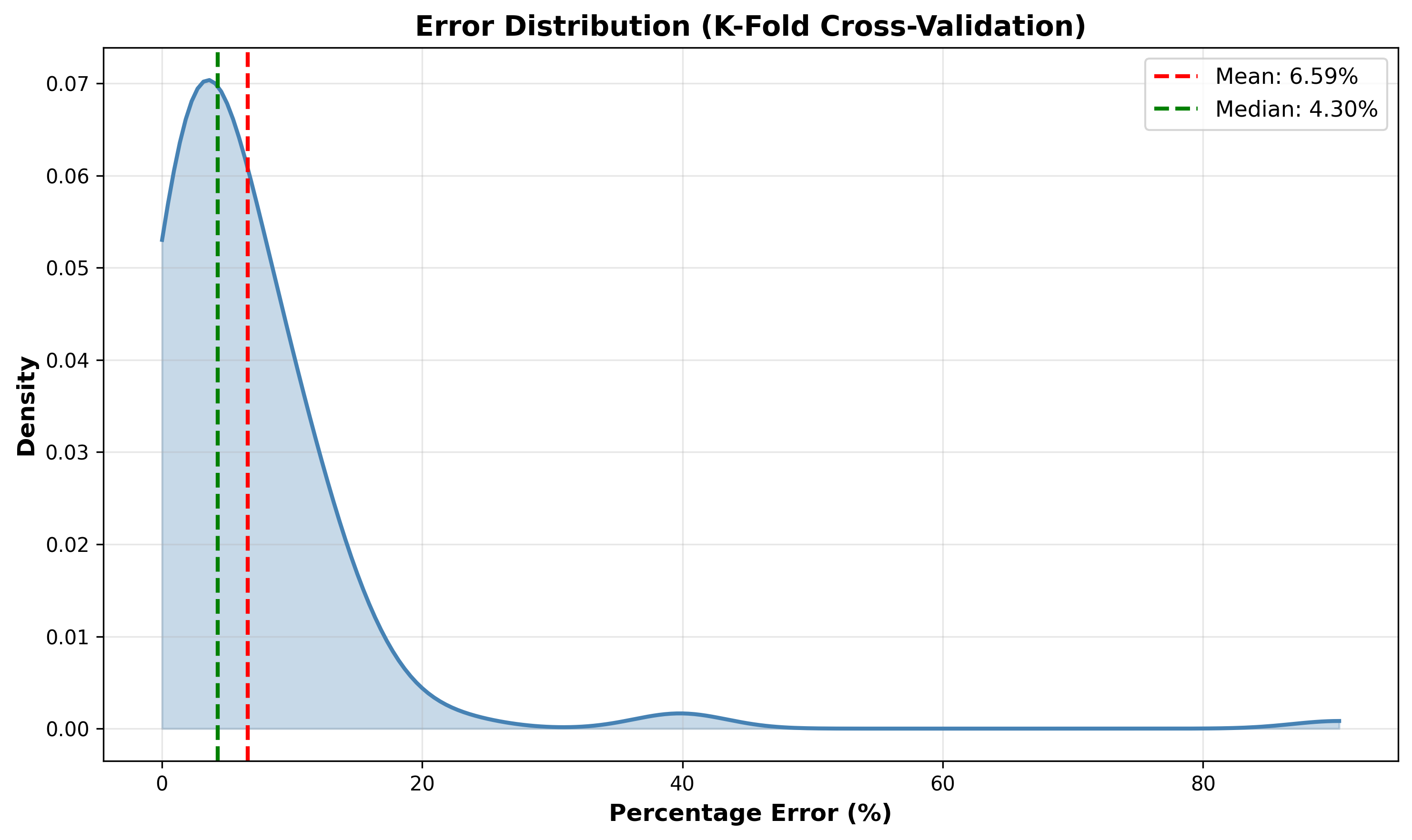}
 \caption{Using informative priors}
    \label{fig:dis_errors}
\end{subfigure}
\begin{subfigure}{0.5\textwidth}
\includegraphics[width=0.9\linewidth]{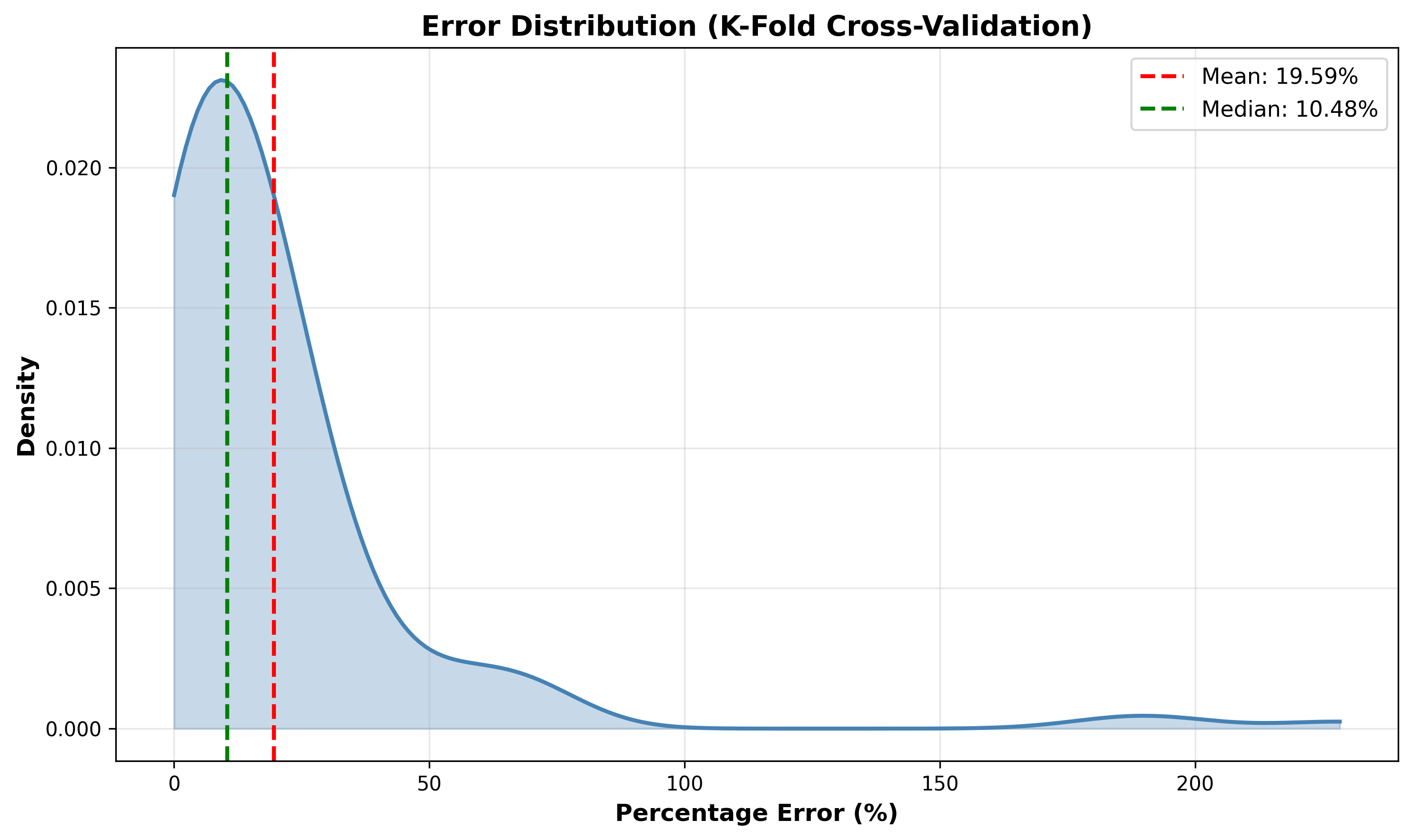}
\caption{Using \cite{Palacios_2014} non-informative priors}
    \label{fig:sid_errors_no_inf}
\end{subfigure}

\caption{Distribution of absolute percentage errors across all test points in the 5-fold cross-validation}
\label{fig:dis}
\end{figure*}



Several factors may explain this performance discrepancy:

\section{Setup} \cite{Palacios_2014} utilizes multiple Markov chains with random initial values, whereas our current implementation is based on a single chain. We initially aim for a ``Keep It Simple'' approach to facilitate understanding of the impact of each parameter.

\section{Dataset size} The dataset in ~\cite{Palacios_2014} is substantially larger ($\approx$ 100x) than the one used in this work. As previously noted, experiments in fermentation environments are slow and resource-intensive. The encouraging results reported in this work provide strong motivation for collecting larger datasets.

\section{Measurement scale} Both studies evaluate the growth measuring the optical density (OD), but at different wavelengths (595~nm in ~\cite{Palacios_2014} versus 600~nm in this work) and potentially different concentration ranges. The MSE is scale-dependent; if the observed OD values in this work span a larger range, the MSE will naturally be higher even if the relative prediction accuracy is comparable: unfortunately, we don't know the initial bacterial concentrations used in ~\cite{Palacios_2014} experiments, but this could be an additional motivation for this discrepancy.

\section{Phenomenon complexity} The environmental parameters studied in ~\cite{Palacios_2014}  (temperature, pH, and NaCl concentration) are well-characterized in the microbiological literature and exhibit relatively smooth, monotonic effects on the growth. In contrast, our work investigates ultrasonic irradiation effects (frequency and duty cycle), which may induce less predictable nonlinear responses and complex interactions.

\begin{figure}[htb]
    \centering
    \includegraphics[width=1\linewidth]{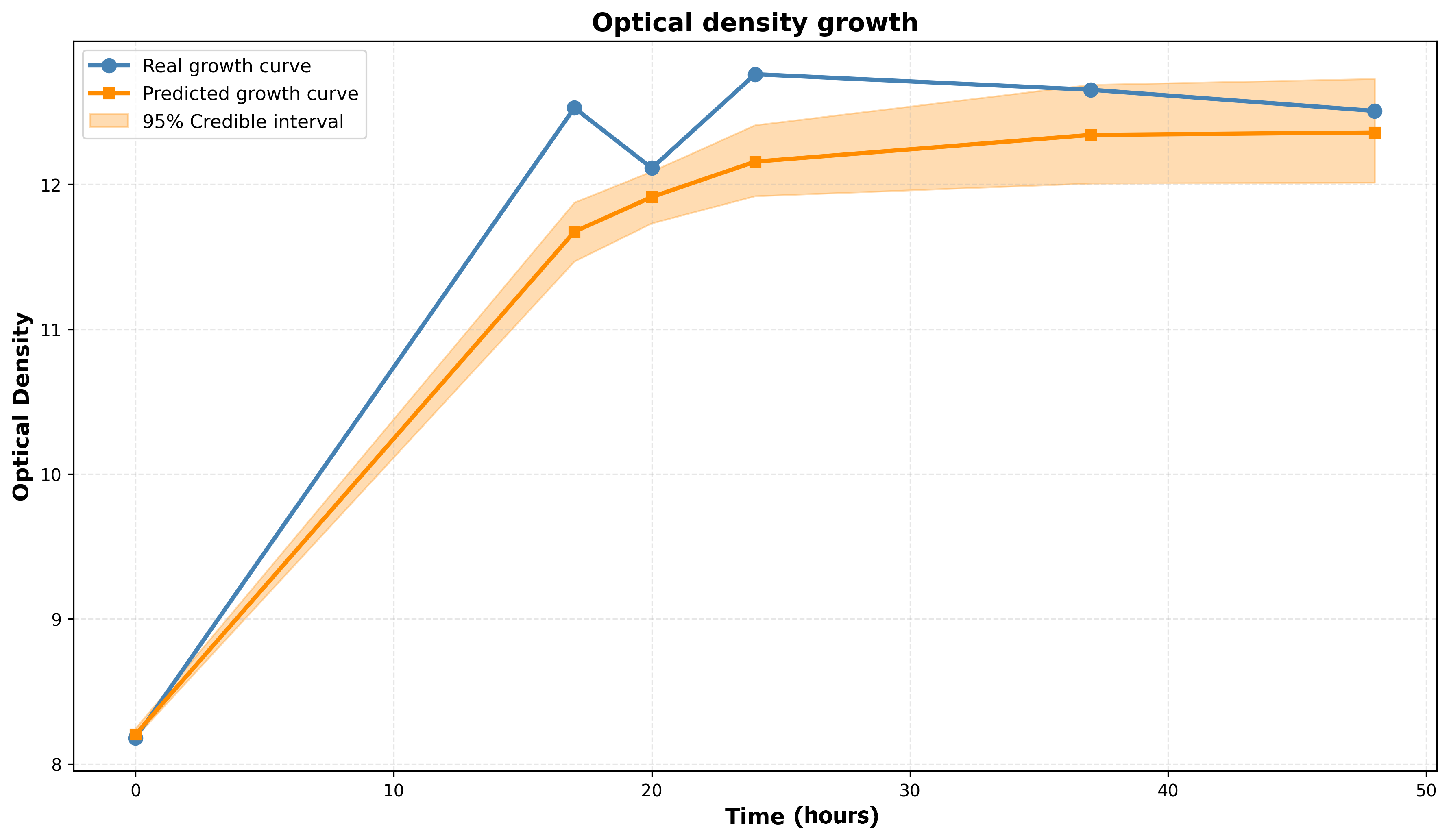}
    \caption{Real growth curve (in blue) expressed as absorbance at 600 nm of the yeast culture compared with the model prediction (in bright orange) and associated confidence interval (orange shaded area). The specific environmental conditions are omitted for confidentiality reasons.}
    \label{fig:growth_curve_comparison}
\end{figure}

\section{Prior Knowledge Effects}

Figure~\ref{fig:errors_no_inf} shows the outcomes of the same experiments of Figure~\ref{fig:errors}, namely predicted versus observed optical density values, when  the non-informative priors reported in~\cite{Palacios_2014} are used. The increased scatter compared to figure~\ref{fig:errors} demonstrates the importance of priors for limited datasets. This is also confirmed by the fact that the MSE goes from $\approx0.08$ to $\approx0.63$. The corresponding distributions of errors are shown in Figure~\ref {fig:dis}.

\chapter{Act}

The ultimate goal of the digital twin, for which this paper lays the foundations, is to enable automated actions that optimize the beer fermentation process based on its predictions. Accelerating yeast growth can shorten the overall production cycle, resulting in significant economic benefits. To this end, the digital twin controls the ultrasonic stimulation parameters, namely frequency, duty cycle, and irradiation timing, while temperature regulation, being a more established task, is left for future work.

The frequency was irradiated using piezoelectric transducers (the ones employed for example in ultrasonic cleaning tanks). These transducers were driven by applying their nominal resonance frequency through a square wave generated via PWM by a microcontroller and amplified using an H-bridge circuit. Duty cycle control was achieved by delivering energy bursts to the transducer rather than a continuous resonance signal over time. These packets carrying the resonance frequency were modulated at a low frequency (150~Hz), although higher modulation frequencies could be used to avoid acoustic noise in the human audible range, provided that the carrier frequency remains lower than the frequency it carries.

\chapter{Conclusion and Future Work}

This work presents a complete digital-twin pipeline for an innovative ultrasonic fermentation system, combining real-time sensing, Bayesian yeast-growth modeling, and intelligent actuation. The system establishes the foundation for autonomous experimentation and optimization in bio-oriented intelligent environments. One evident advantage is represented Time consuming real experiments to explore the phase space of variables could be complemented and instructed by the quicker and resource-friendly simulations of the digital twin counterpart. 

Despite the higher MSE compared to the reference study~\cite{Palacios_2014}, the causes of which are discussed in Section~\ref{sec:predict}, the results shown in Figure~\ref{fig:growth_curve_comparison} exhibit strong qualitative agreement, providing an initial validation of the model.
Nevertheless, a substantial expansion of the dataset, coupled with a relaxation of the prior knowledge constraints in the model, is required to strengthen the results, and is one of the aims for the follow-up studies. This data-collection effort aligns with the iterative, continuous-improvement philosophy of the digital twin framework, in which sensors continuously gather data to refine the predictive model.

Future work includes thermal control and the integration of automated optical-density sensing, currently cost-prohibitive, to fully close the feedback loop.
\\

\bibliographystyle{plain}
\bibliography{bibliography/biblio}

\end{document}